\documentclass[runningheads]{llncs}
\usepackage[T1]{fontenc}
\usepackage{graphicx}
\usepackage{caption}
\usepackage{subcaption}
\usepackage{adjustbox}
\usepackage[backref=page]{hyperref}
\usepackage{orcidlink}
\usepackage{xcolor}
\hypersetup{
    colorlinks=true,
    linkcolor=blue,
    urlcolor=blue,
    citecolor=black,
    filecolor=black,
    linkbordercolor=white, 
    urlbordercolor=white,
    citebordercolor=white,
    filebordercolor=white,
}

\usepackage{hhline}
\usepackage{multirow}
\usepackage{makecell}
\usepackage{comment}

\usepackage{siunitx}

\usepackage{float} 
\usepackage[noabbrev]{cleveref}

\usepackage{color}

\usepackage[misc]{ifsym}

\begin{document}
\title{\texorpdfstring{From FDG to PSMA: A Hitchhiker's Guide\\to Multitracer, Multicenter Lesion Segmentation\\in PET/CT Imaging}{From FDG to PSMA: A Hitchhiker's Guide to Multitracer, Multicenter Lesion Segmentation in PET/CT Imaging}}
%
%
\author{Maximilian Rokuss\inst{1,3}\orcidlink{0009-0004-4560-0760} \and 
        Balint Kovacs\inst{1,2}\orcidlink{0000-0002-1191-0646} \and
        Yannick Kirchhoff\inst{1,3}\orcidlink{0000-0001-8124-8435} \and\\
        Shuhan Xiao\inst{1,3} \and
        Constantin Ulrich\inst{1,2}\orcidlink{0000-0003-3002-8170} \and\\
        Klaus H. Maier-Hein\inst{1,4,5,\star}\orcidlink{0000-0002-6626-2463}\and
        Fabian Isensee\inst{1,4,\star}\orcidlink{0000-0002-3519-5886}
}
\authorrunning{M. Rokuss et al.}
\titlerunning{DKFZ tackles autoPET III}

\institute{
German Cancer Research Center (DKFZ) Heidelberg,\\Division of Medical Image Computing, Heidelberg, Germany
\and
Medical Faculty Heidelberg, Heidelberg University, Heidelberg, Germany\and
Faculty of Mathematics and Computer Science,\\Heidelberg University, Heidelberg, Germany\and
Helmholtz Imaging, DKFZ, Heidelberg, Germany\and
Pattern Analysis and Learning Group, Department of Radiation Oncology, Heidelberg University Hospital, Heidelberg, Germany
\email{maximilian.rokuss@dkfz-heidelberg.de}
}

\maketitle              

\renewcommand{\thefootnote}{$\star$}
\footnotetext[1]{Shared Last Authorship}

\begin{abstract}
Automated lesion segmentation in PET/CT scans is crucial for improving clinical workflows and advancing cancer diagnostics. However, the task is challenging due to physiological variability, different tracers used in PET imaging, and diverse imaging protocols across medical centers. To address this, the autoPET series was created to challenge researchers to develop algorithms that generalize across diverse PET/CT environments. This paper presents our solution for the autoPET III challenge, targeting multitracer, multicenter generalization using the nnU-Net framework with the ResEncL architecture. Key techniques include misalignment data augmentation and multi-modal pretraining across CT, MR, and PET datasets to provide an initial anatomical understanding. We incorporate organ supervision as a multitask approach, enabling the model to distinguish between physiological uptake and tracer-specific patterns, which is particularly beneficial in cases where no lesions are present. Compared to the default nnU-Net, which achieved a Dice score of 57.61, or the larger ResEncL (65.31) our model significantly improved performance with a Dice score of 68.40, alongside a reduction in false positive (FPvol: 7.82) and false negative (FNvol: 10.35) volumes. These results underscore the effectiveness of combining advanced network design, augmentation, pretraining, and multitask learning for PET/CT lesion segmentation. After evaluation on the test set, our approach was awarded the first place in the model-centric category (\textit{Team LesionTracer}). Code is publicly available at \url{https://github.com/MIC-DKFZ/autopet-3-submission}.

\keywords{PET/CT \and Lesion Segmentation \and Generalization.}
\end{abstract}
\renewcommand\thefootnote{\arabic{footnote}}
\setcounter{footnote}{0}
\section{Introduction}
Positron Emission Tomography (PET) combined with Computed Tomography (CT) is a powerful tool in modern medical diagnostics, particularly for detecting and monitoring cancer. PET/CT scans provide both metabolic and anatomical information, allowing clinicians to identify tumor lesions with high precision. However, manual segmentation of lesions in PET/CT images is time-consuming, and labor-intensive rendering it infeasible for patients with a multitude of lesions. Automated lesion segmentation offers a promising solution, enabling faster, more consistent analysis, which is crucial for clinical workflows and research.\\

\noindent However, despite its potential, automated segmentation faces significant challenges. The complexity arises from the physiological variability between patients, the different tracers used in PET imaging (such as FDG and PSMA), and the variations in imaging protocols across different medical centers. Each tracer can highlight different metabolic activities, which leads to distinct patterns of uptake in non-tumor structures, complicating the task of distinguishing between normal physiological uptake and actual lesions. To accurately assess PET/CT images, a model must learn to interpret varying uptake patterns without explicit information about the specific tracer used. Instead, it must rely on the surrounding anatomical context to differentiate between physiological and cancerous uptake. These complexities, particularly the variability in tracer behavior, have made automated lesion segmentation in PET/CT imaging a highly challenging task for models to perform effectively.\\

\noindent To address this, the autoPET challenge series was created, offering a platform for researchers to directly tackle these issues. Building on the insights from previous iterations, the \href{https://autopet-iii.grand-challenge.org}{autoPET III challenge} broadens its scope to focus on multitracer, multicenter generalization.  The publicly available dataset of 1014 FDG PET/CT studies \cite{gatidis2020fdgpetct} has been extended by 597 exams with a new PSMA tracer \cite{jeblick2024psmapetct}. By providing access to large, annotated datasets from different hospitals, participants are tasked with developing algorithms capable of accurate and robust segmentation across diverse PET/CT environments. This challenge represents a crucial step toward enhancing automated medical imaging for real-world clinical applications.\\

\noindent This manuscript describes our participation in the \href{https://autopet-iii.grand-challenge.org}{autoPET III challenge} which later on won the first place in the model-centric category. We base our solution on a strong foundation offered by nnU-Net\cite{isensee2021nnu} and address the aforementioned challenges of automated PET/CT lesion segmentation through data augmentation, pretraining, model design, and postprocessing techniques. Our approach aims to improve generalization across tracers and centers, tackling the complexities of physiological and cancerous uptake.

\section{Methods}
\label{sect:methods}
Our method builds on the well-established nnU-Net framework\cite{isensee2021nnu}, specifically, we opt for a larger and more powerful network given by the recently introduced ResEncL architecture preset \cite{isensee2024nnu}.

\subsection{nnUNet Configuration}
We use the \verb+3d_fullres+ configuration, resample all images to a common spacing of \verb+[3, 2.04, 2.04]+ and normalize both modalities with the default CT normalization scheme. We train with a batch size of 2 for 1000 epochs and a uniform patch size of \verb+192x192x192+ for all trainings during method development. This large patch size has the advantage of providing the network with more context which is important, especially for the task at hand where the network needs to infer the tracer type and uptake rate from the surrounding organs. We also noticed that training without a smoothing term in the dice loss calculation the training becomes more stable, hence we omit it. The best model we use for the submission is retrained with a batch size of 3 for 1500 epochs.

\subsection{Data augmentation}
\label{subsec:DA}
To account for potential misalignments \cite{alessio2004pet,hunter2016patient,kaji2024improvement,lodge2011effect} between the CT and PET images, we extended the data augmentation (DA) scheme of nnU-Net with misalignment DA \cite{kovacs2023addressing}. Essentially, this augmentation shifts the PET and CT images relative to each other to make the network more robust to incorrect spatial alignment. This approach has the potential advantage of improving sensitivity for punctate lesions with small voxel segmentation volumes, which was indicated in the original study but remained unproven. The amplitude of the transformations used to generate misalignments was sampled randomly from a uniform distribution, constrained by a maximum amplitude in both positive and negative directions. The transformation included an initial rotation with a maximum angle of \SI{5}{\degree}, followed by translations with maximum voxel shifts of \verb+[2, 2, 0]+ in the x, y, and z directions, respectively.

\subsection{Pretaining and Finetuning}\label{subsec:training}

To steer the model towards an anatomically relevant loss minimum, we first pretrain it on a large and diverse dataset of 3D medical images, combining a variety of public datasets in a MultiTalent-inspired fashion \cite{ulrich2023multitalent}. Initially restricted to CT datasets, we later expanded this pretraining to include datasets from PET and Magnetic Resonance Imaging (MRI) modalities, allowing the model to learn general features and develop a universal understanding of anatomy and medical images. Employing separate segmentation heads for each dataset, the model was trained for 4000 epochs with a patch size of \verb+[192,192,192]+ and a batch size of 24, resampling all images to a cubic 1mm resolution and Z-score normalization. Dataset sampling was performed inversely proportional to the square root of the number of images per dataset, ensuring balanced training. The resulting foundation model, implemented using the nnU-Net framework and based on the ResEncL U-Net architecture \cite{isensee2024nnu}, can effectively segment multiple datasets simultaneously and demonstrates strong potential for fine-tuning on specific downstream tasks. We have made the pretrained model weights publicly available to encourage further research\footnote{\url{https://zenodo.org/records/13753413}}. After this pretraining, we fine-tune the model on the autoPET III dataset with the initial learning rate lowered to 1e-3, enhancing robustness and segmentation accuracy by leveraging weights optimized on a broad range of tasks.

\subsection{Organ Supervision}
\label{subsec:organ_supervision}

\begin{figure}[t]
    \centering
    \begin{subfigure}[t]{0.32\textwidth}
        \centering\includegraphics[width=\textwidth]{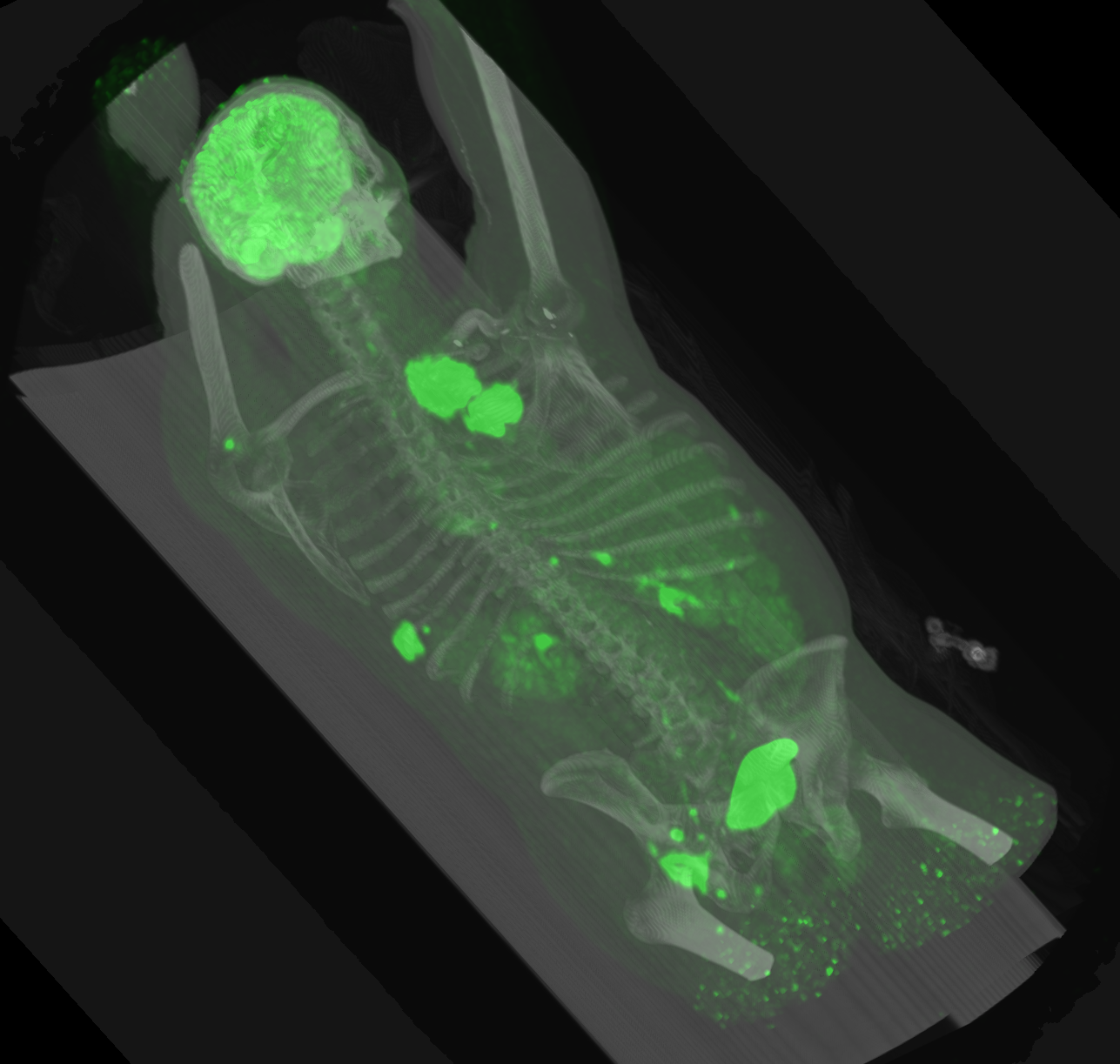}
        \caption{PET/CT Image}
    \end{subfigure}
    \begin{subfigure}[t]{0.32\textwidth}
        \centering\includegraphics[width=\textwidth]{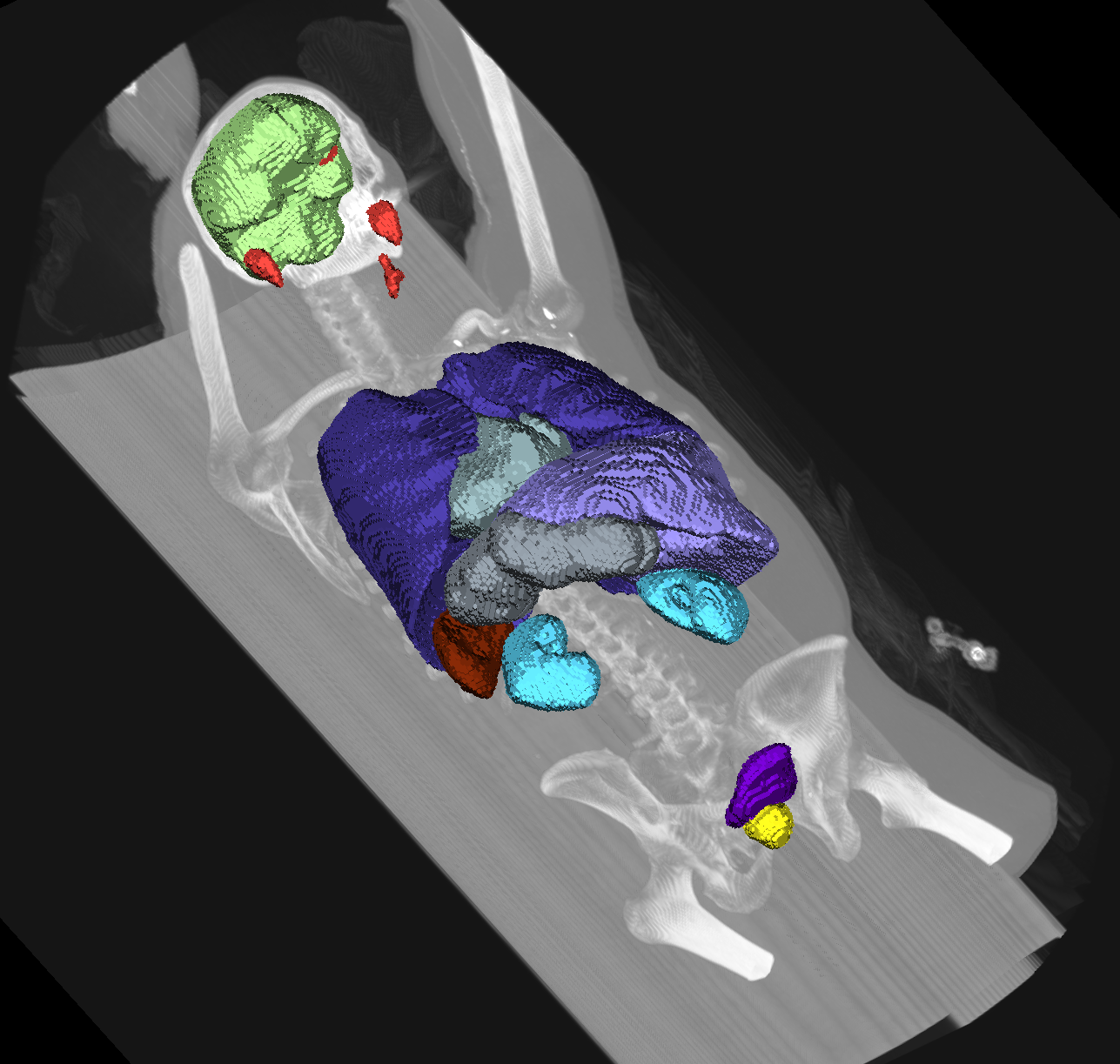}
        \caption{Organ annotation}
    \end{subfigure}
    \begin{subfigure}[t]{0.32\textwidth}
        \centering\includegraphics[width=\textwidth]{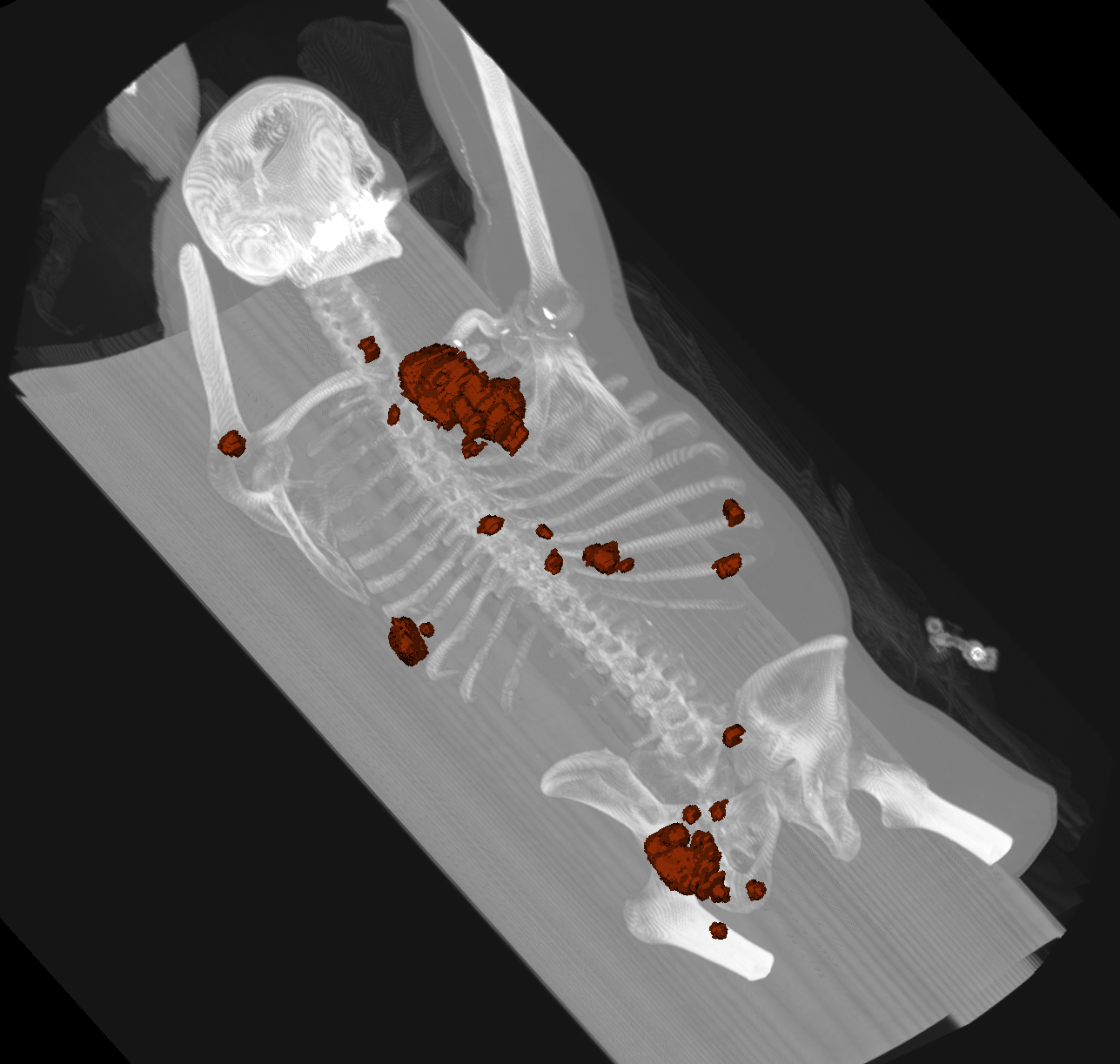}
        \caption{Lesion annotation}
    \end{subfigure}
    \caption{Examplary visualization of case \texttt{fdg\symbol{95}0b98dbe00d\symbol{95}08-11-2002} shows the FDG PET image overlaid on the CT scan (a), along with corresponding annotations for organs (b) and lesions (c), used in the multitask training process.}
    \label{fig:organs}
\end{figure}

To improve further anatomical understanding and enhance segmentation performance, we introduce an additional prediction head focused on segmenting key organs. These organs often exhibit higher tracer uptake, which may not be the primary target but could complicate diagnosis. By addressing these areas, the network aims to reduce false positive volume (FPvol) and improve overall accuracy. We use TotalSegmentator \cite{wasserthal2023totalsegmentator,totalseg_glands} to predict the spleen, kidneys, liver, urinary bladder, lung, brain, heart, stomach, prostate, and glands in the head region (parotid glands, submandibular glands) for all images as shown in Fig. \ref{fig:organs}. The selection of these structures was influenced by the uptake patterns of various tracers, reflecting the differing behaviors observed between them. This addition complements the lesion segmentation head, resulting in a dual-headed architecture where one head focuses on lesions and the other on organ structures. Each prediction head is then trained using a softmax activation function and with equal loss weighting during optimization.\\

\noindent The rationale behind organ supervision is to provide additional guidance to the network by introducing anatomical priors and focusing on organs that exhibit uptake for specific tracers. This is particularly important given that approximately half of the training cases do not contain lesions, limiting the direct lesion-based learning signal. By including organ supervision, we ensure that the network learns meaningful representations of both healthy and abnormal anatomical structures, promoting better generalization. This approach helps improve segmentation performance, especially in cases without lesions, where the network can rely on organ structures for guidance. We also observed a faster convergence of lesion segmentation during training with the addition of organ supervision.\\

\noindent We also experimented with a variant where the organ supervision loss was assigned a lower weight relative to the lesion segmentation loss. While this variant showed a slight improvement in Lesion Dice scores, it led to an increase in the false positive and false negative volume, suggesting a trade-off in performance. Consequently, we found that balancing the losses equally for both prediction heads yielded the most robust overall results.

\subsection{Approaches we tried, that didn't work}

We also explored several ablation strategies that did not yield significant improvements during development. Resampling the images to an isotropic spacing of \verb+[1, 1, 1]+ resulted in a substantial performance decline, likely due to the reduced contextual information available within each patch. Additionally, pretraining on the TotalSegmentator dataset \cite{wasserthal2023totalsegmentator} provided no notable advantage over training the model from scratch. Similarly, incorporating additional annotated FDG images from the HECKTOR challenge \cite{hecktor} into the training set did not enhance the performance metrics on the autoPET dataset.

\section{Results}

\begin{table}[t]
        \begin{center}
        \caption{Five-fold cross-validation results. The table shows the metrics calculated in the official evaluation as well as the Dice values for the FDG and PSMA tracer separately.}
        \label{tab:results}
            \adjustbox{max width=\textwidth}{%
                \begin{tabular}{l|ccc|cc}
                & \multicolumn{3}{c|}{Dice$\uparrow$} & & \\
            Setting & All & FDG & PSMA  & FPvol$\downarrow$ & FNvol$\downarrow$  \\ \hline
            nnU-Net & 57.61 & 63.93    & 51.69     &          19.32      & 15.69 \\
            nnU-Net ResEnc L & 65.31 & 72.87    & 58.25     &  10.47      & 13.63 \\  
            + misalDA  & 65.76 & 73.13    & 58.89     &        10.12     & 12.50 \\
            + pretrain CT & 66.08 & 73.13    & 59.46     &  7.82      & 15.34 \\
            + pretrain CT, misalDA & 66.37 & 73.06  & 59.99 & 7.99  & 14.46 \\
            
            + pretrain CT, noSmooth & 67.01 & 75.04  & 59.50 & 10.08  & 13.50 \\
            + pretrain CT, misalDA, noSmooth & 67.45 & 75.41  & 60.01 & 9.45  & 13.33 \\
            + pretrain CT/MR/PET,  misalDA, noSmooth & 68.03 & 75.95  & 60.63 & 9.39  & 13.52 \\
            + pretrain CT/MR/PET, misalDA, noSmooth, organs & 68.33 & 76.30  & \textbf{60.84} & 8.93  & \textbf{10.15} \\ \hline
            + batch size 3, 1500 epochs  & \textbf{68.40} & \textbf{77.28}  & 60.01 & \textbf{7.82}  & 10.35 \\
            \end{tabular}
            }
        \end{center}
    \end{table}

Model development and evaluation were carried out using a five-fold cross-validation on the autoPET III training split. As shown in Table \ref{tab:results}, the nnU-Net ResEnc L architecture outperformed the baseline nnU-Net across all metrics, serving as the backbone for further experimentation.\\

\noindent Initial improvements focused on data augmentation and pretraining strategies. Incorporating the misalignment data augmentation (misalDA) resulted in a marginal improvement in Dice score (from 65.31 to 65.76), particularly benefiting the PSMA tracer and achieving a low FNvol. Supervised pretraining the model on CT datasets boosted performance further, especially for PSMA, increasing the overall Dice to 66.08.\\

\noindent Combining CT pretraining with misalDA led to a more balanced improvement across both tracers compared to the ResEncL baseline, pushing the Dice to 66.37. Additional experiments with no smoothing term in the dice calculation yielded a significant leap, especially for FDG cases, with the overall Dice reaching 67.45. \\

\noindent Another gain was achieved by pretraining on multimodal datasets (CT/MR/PET) paired with misalDA. Lastly, the inclusion of organ segmentation as a multitask learning objective further enhanced the model, achieving a Dice score of 68.33 and notably reducing both false positive and false negative volumes.  Retraining this model with a batch size of 3 for 1500 epochs yields the best overall model with a Dice score of 68.40 and the lowest combined FP and FN volume. 

\subsection{Test Set Submission}

For the final submission, we ensembled all 5 folds. To meet the 5-minute time constraint per case, we set a tile step size of 0.6, which controls the sliding window shift relative to the patch size. The first fold is processed without mirroring, and based on the time taken, one or two mirroring axes are added as test-time augmentation for the remaining folds.

\section{Conclusion}

In this paper, we addressed the challenges of automated lesion segmentation in PET/CT imaging through a combination of data augmentation, pretraining, and multitask learning as well as a careful choice of the underlying network design and training. Building on the nnU-Net framework, we use misalignment data augmentation and multimodal pretraining, which improved generalization across different tracers and centers. Incorporating organ supervision as a secondary task further boosted performance by guiding the model with anatomical priors, especially in cases without lesions. This holistic approach resulted in significant improvements, achieving a top Dice score of 68.33 on the training set cross-validation, demonstrating the potential of these techniques in advancing multitracer, multicenter lesion segmentation.
    
\begin{credits}
\subsubsection{\ackname}
Part of this work was funded by Helmholtz Imaging (HI), a platform of the Helmholtz Incubator on Information and Data Science. The present contribution is supported by the Helmholtz Association under the joint research school "HIDSS4Health – Helmholtz Information and Data Science School for Health".
\end{credits}

\newpage
\bibliographystyle{splncs04}
\bibliography{biblio1}

\end{document}